\documentclass[12pt]{emulateapj}
\usepackage{graphicx}
\usepackage[bookmarks=false]{hyperref} 

\newcommand{\swift}{\emph{Swift}}
\newcommand{\fermi}{\emph{Fermi}}

\newcommand{\e}[1]{10$^{#1}$}
\newcommand{\ee}[1]{$\times$10$^{#1}$}

\shorttitle{Precursors of short GRBs}
\shortauthors{E. Troja et al.}

\begin{document}
\submitted{Submitted 2010 June 7; accepted 2010 September 7}
\title{Precursors of short gamma-ray bursts}

\author{E. Troja\altaffilmark{1,2}, S. Rosswog\altaffilmark{3},
N. Gehrels\altaffilmark{1}
}

\altaffiltext{1}{NASA, Goddard Space Flight Center, Greenbelt, MD 20771, USA}
\altaffiltext{2}{NASA Postdoctoral Program Fellow}
\altaffiltext{3}{School of Engineering and Science, Jacobs University Bremen, 
Campus Ring 1, 28759 Bremen, Germany}

\begin{abstract}

We carried out a systematic search of precursors on the sample
of short GRBs observed by \swift. We found that $\sim$8-10\% of short GRBs
display such early episode of emission. One burst (GRB~090510) 
shows two precursor events,  the former $\sim$13~s and the 
latter $\sim$0.5~s before the GRB.
We did not find any substantial difference between the precursor
and the main GRB emission, and between short GRBs with
and without precursors. 
We discuss possible mechanisms to reproduce the observed precursor
emission within the scenario of compact object mergers.
The implications of our results on quantum gravity constraints 
are also discussed. 

\end{abstract}


\section{Introduction}\label{sec:intro}

The main gamma-ray event in GRBs is occasionally anticipated
by a less intense episode of emission, called a precursor. 
With the exception of a few cases, precursors show non-thermal spectra
and have been mainly observed in long duration GRBs \citep[e.g.][]{l05, b08}.  
The first evidence of  preburst activity has been observed by the Ginga satellite in 
the long GRB~900126 \citep{murakami91}, where a soft X-ray peak 
precedes the burst onset by $\sim$8 s. This is one of the
few examples of a precursor with a thermal spectrum.

Observationally, the identification of a precursor 
is highly dependent on its operational definition
and might further suffer of instrumental biases.
A first systematic search of precursors \citep{k95}
showed that only 3\% of GRBs observed by BATSE
exhibits a precursor event, having no substantially different properties
with respect to the $\gamma$-ray prompt emission. 
By using a different search criterion, 
\citet{l05} found instead that $\sim$20\% of long GRBs 
are preceded by an early emission episode, 
which is much weaker and spectrally softer than the proper GRB.
This result was also confirmed by Beppo-SAX \citep{piro05} 
and HETE-2 \citep{hete04} observations, yet it is unclear 
whether it still holds for the precursor activity detected in \swift~GRBs. 
Hints of different properties (e.g. hardness ratio, spectral lag) 
between the precursor and the prompt emission
have been reported in the study of single bursts \citep[e.g. GRB061121;][]{page07}.
However the recent work of \citet{b08}, based on a large sample of long GRBs observed 
by \swift, did not find evidence for such spectral distinction. 

An ubiquitous feature, emerging from all the previous studies, is the 
distribution of delay times between the precursor and the prompt emission,
which extends up to hundreds of seconds. 
This represents one of the main challenges to the current theoretical models.
If the precursor marks the start of the central engine activity 
\citep{nakamura00}, 
the observed quiescent time would require a fine tuning of the 
ejecta Lorentz factors 
or an effective turn-off of the GRB energy source \citep{err01}.
However, whether this early emission  physically differs from the burst itself 
or is part of the same event remains a controversial point.  
Precursors as a separate phenomenon have also been discussed in several 
theoretical scenarios \citep[e.g.][]{lyutikov00,merees01,waxman03,lyutikov03,umeda05,wang07}. 
A set of models explain the precursor emission within the 
standard fireball scenario, 
commonly invoked to interpret the prompt and afterglow emission of GRBs. 
Within this framework the precursor is associated
with the transition of the fireball to the optically thin regime, 
which produces a photospheric blackbody emission \citep{pac86,merees01,daigne02,ruffini08}, 
while the GRB is associated
with the later formation of shocks at larger radii \citep{remes94}. 
According to this interpretation, precursors 
happen relatively close in time to the GRB and 
should be observable in both long and short bursts. 

Another class of models interpret the precursor
within the collapsar scenario and link its origin  
to the jet break-out from the stellar surface 
\citep[e.g.][]{err02,waxman03,zhangw03,lb05}.
The temporal delay between the precursor and the GRB 
is explained as an apparent period of quiescence
due to viewing angle effects \citep[e.g.][]{morsony07}
or, alternatively, as an intrinsic property of the central engine,
which might undergo a second collapse \citep[][]{wang07}.
In this case, precursors might be observed tens of seconds
before the GRB emission and
should occur exclusively in long GRBs.
Indeed most of the observational
and theoretical effort so far has been focused on long duration GRBs.

Little attention has been paid to the occurrence of precursors
in short GRBs, and the possibility of an early precursor emission 
originated in the last moments of a compact binary merger
has been sporadically discussed in literature \citep[e.g.][]{hansen01,rosswog02b}.
In this paper we explore in detail the observational evidence for precursors
in short GRBs and discuss the implication.
The paper is organized as follows: 
in \S~\ref{sec:data} we describe the selection criteria
and the detection method adopted to identify the precursor emission;
the resulting sample of precursors
and their properties is presented in \S~\ref{sec:res};
in \S~\ref{sec:discuss} we discuss our findings in the framework
of fireball and progenitor precursors (\S~\ref{sec:parent}). 
In \S~\ref{sec:liv} we focus on  the precursor emission 
observed in GRB~090510 and the implications for quantum 
gravity constraints.

\section{Data analysis}\label{sec:data}

Up to January 2010 \swift~detected 38 GRBs classified
as short bursts ($T_{90}\lesssim2$\,s).   
We included in our sample 11 additional GRBs, which belong to the so-called
group of short bursts with extended emission \citep{nb06}. 
As the classification of these bursts
is unclear \citep{bloom08,zhang09}, we considered them as a separate group.

We searched the selected sample of 49 \swift/BAT GRBs
for signal preceding the main gamma-ray event. 
We define precursors as those events which fulfill the following requirements:
1) the peak flux is smaller than the main event;
2) the flux returns to the background level before the start of the main event;
3) the event location in the sky corresponds to the GRB position.

The \swift~data were retrieved from the public archive\footnote{
\href{http://heasarc.gsfc.nasa.gov/docs/swift/archive/}
{http://heasarc.gsfc.nasa.gov/docs/swift/archive/}}
and processed with the standard \swift~analysis software (v3.5)
included in the NASA's HEASARC software (HEASOFT, ver.~6.8)
and the relevant calibration files. Our analysis has been
performed in the 15-150 keV energy band. 

\subsection{Temporal analysis}\label{sec:wavelet}


\begin{figure*}[!ht]
\centering
\includegraphics[angle=270,scale=0.36]{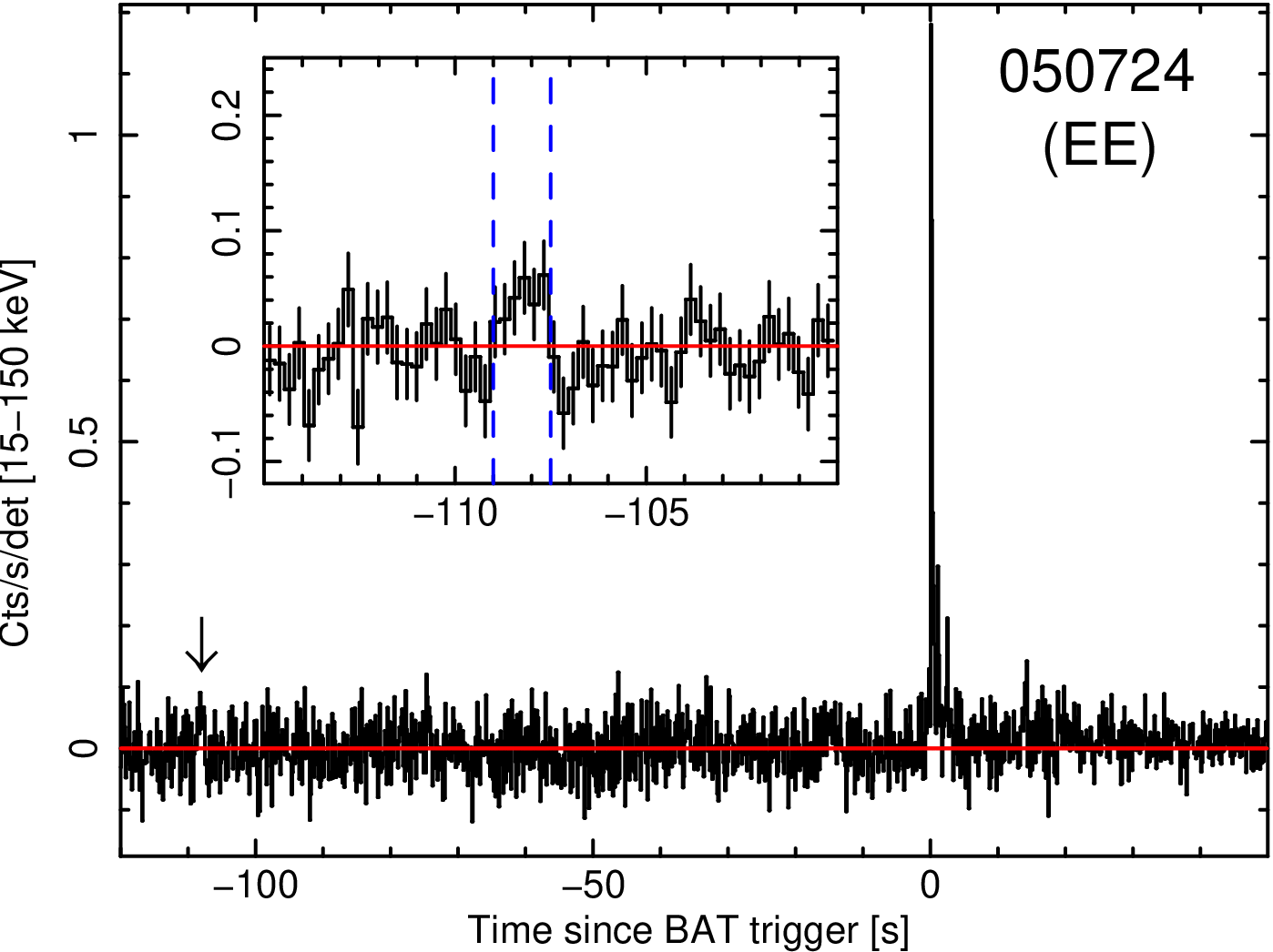}
\hspace{0.8cm}
\includegraphics[angle=270,scale=0.36]{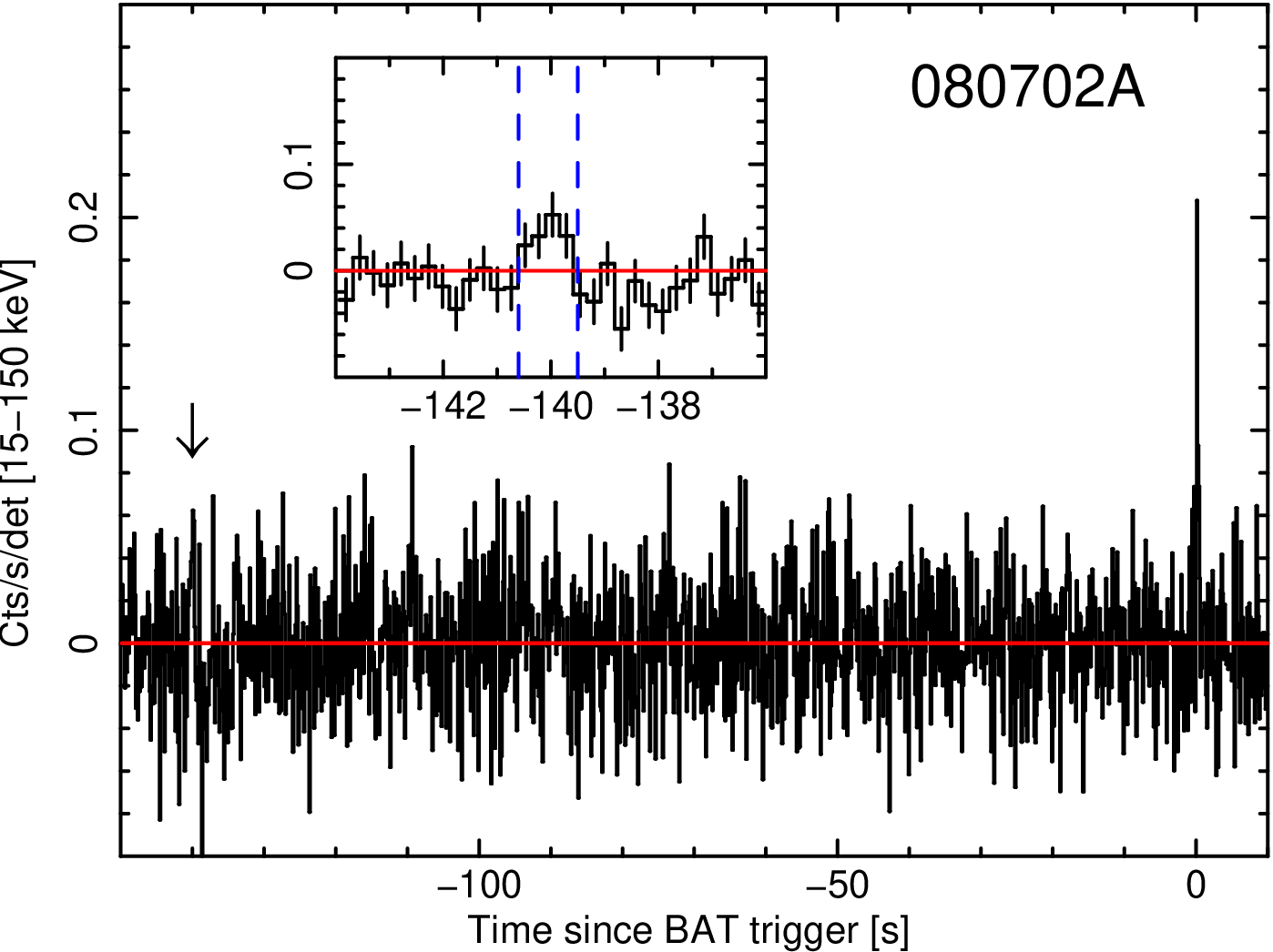}\\
\vspace{0.8cm}
\includegraphics[angle=270,scale=0.36]{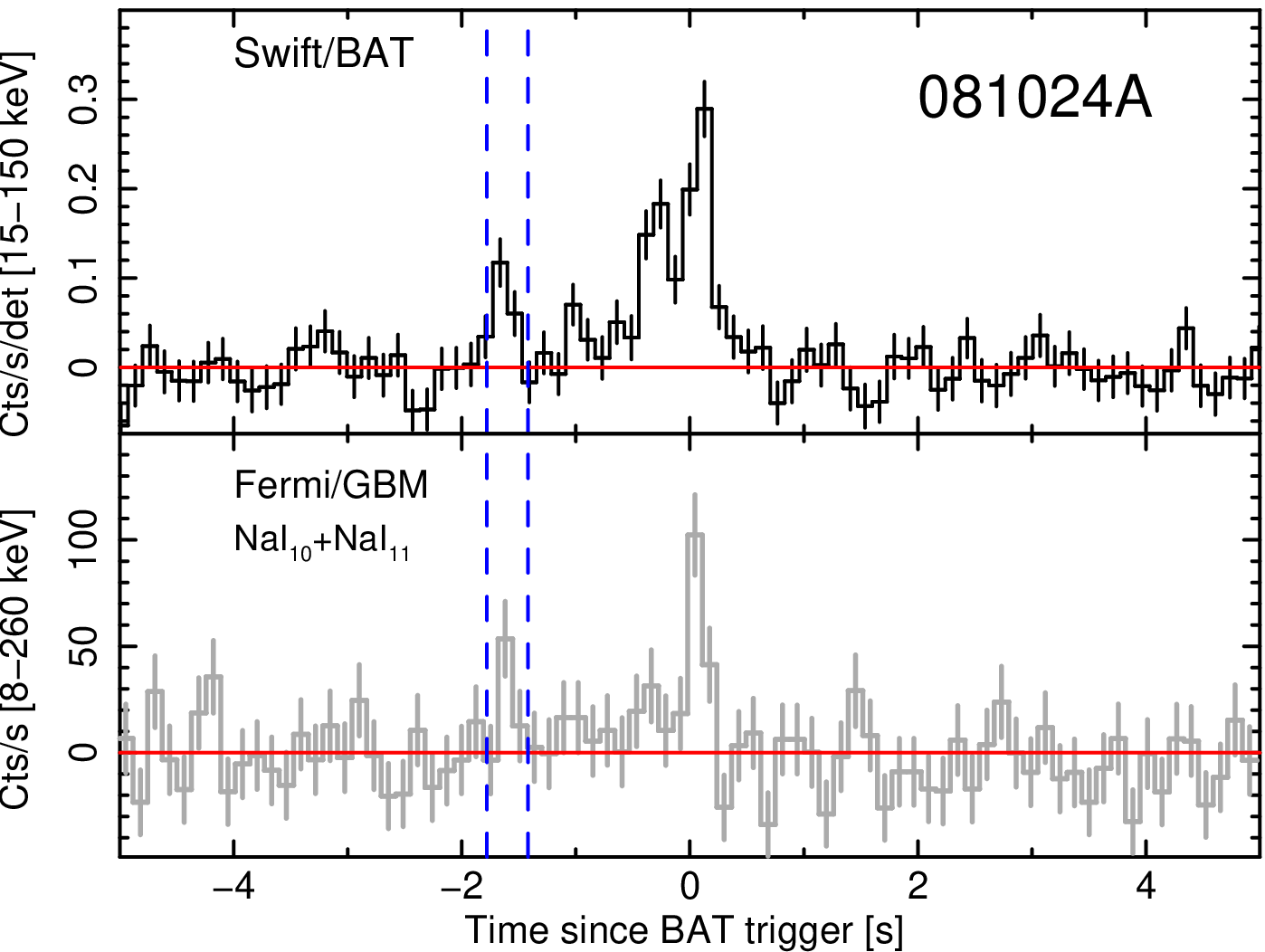}
\hspace{0.8cm}
\includegraphics[angle=270,scale=0.36]{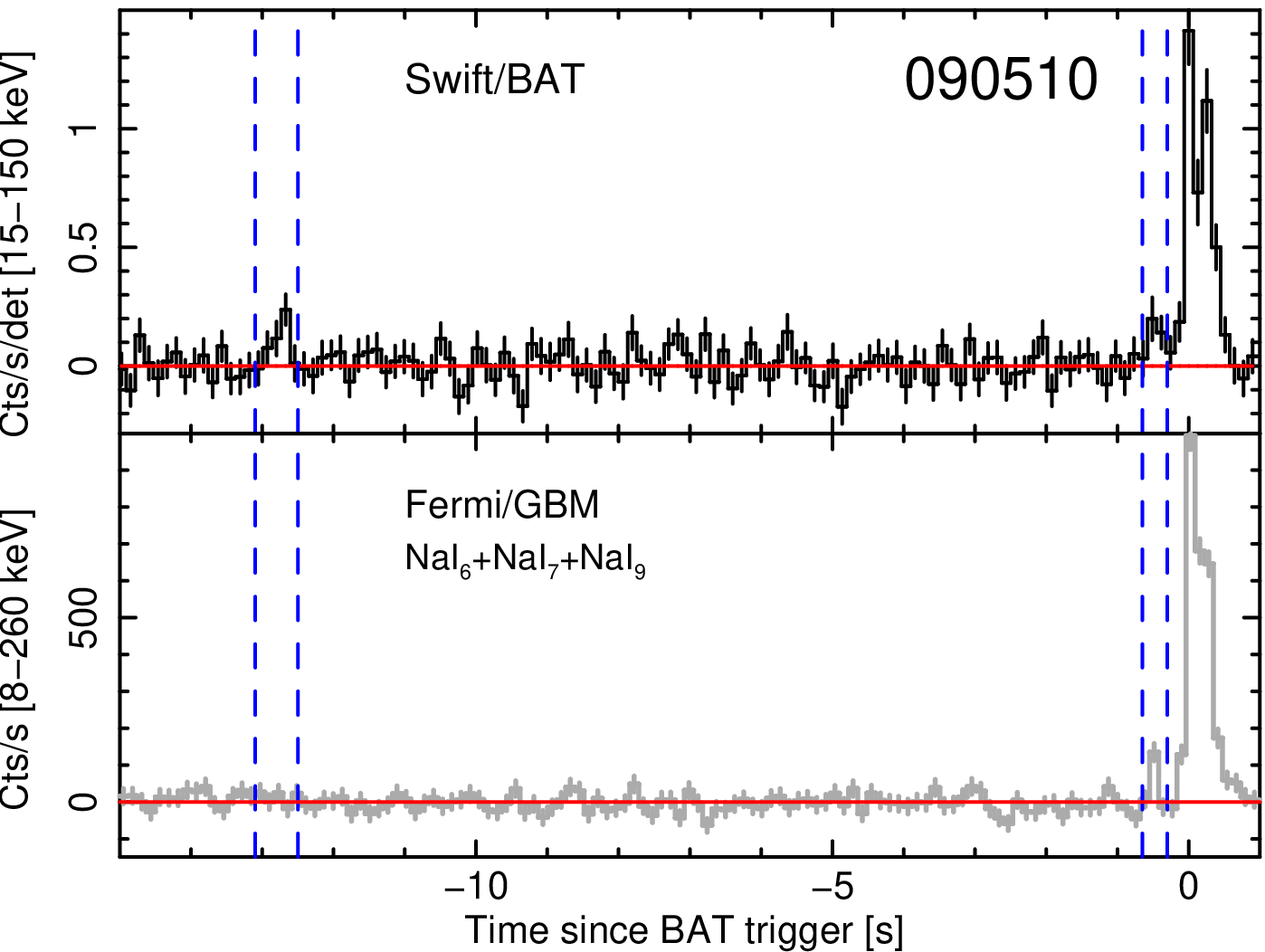}\\
\vspace{0.8cm}
\includegraphics[angle=270,scale=0.36]{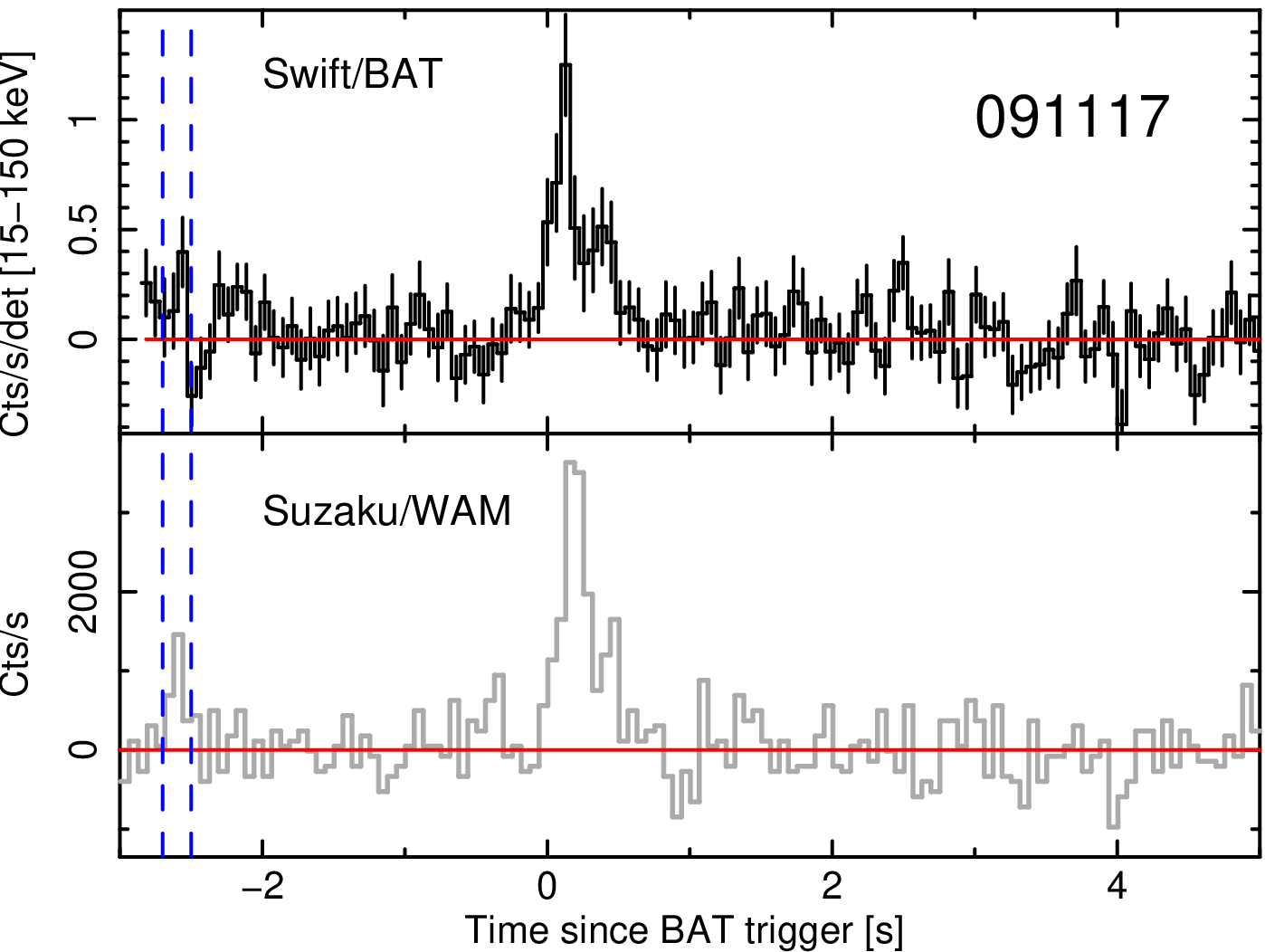}

\caption{\swift/BAT mask-weighted light curves (15--150~keV)
of short GRBs with possible precursor activity. 
Dashed vertical lines mark the precursor duration.
The precursors of GRB080702A and GRB050724 are shown
in greater detail in the insets. 
For comparison, we also show the background-subtracted light curves of 
Fermi/GBM (090510 and 081024A) and Suzaku/WAM (091117). }
\label{fig:batlc}
\end{figure*}


As a first step we inspected the GRB temporal profiles searching for precursor events,
as defined in 1) and 2). 
For each GRB in the sample we created a light curve with a time bin
of 0.128 ms, as shorter time scales are subject 
to noise fluctuations. 
In the case of \swift/BAT on-board triggers $\sim$300 sec of event data 
are usually collected before the GRB trigger.
We excluded from our analysis those time intervals during 
which the spacecraft was slewing, 
starting our search on average $\sim$240\,s before the GRB. 
In order to identify the presence of weak emission in the GRB light curves
we used a detection algorithm based on wavelet transforms \citep{tc98} 
with a Morlet mother function. 
The wavelet algorithm performs a multi scale analysis which is well suited for detecting a transient event, such as a precursor, whose duration is a priori unknown. 
We sampled 13 different time scales with a base-two logarithmic spacing, 
where the smallest resolvable scale $s_0$ is set by the light curve temporal resolution $\delta t$
($s_0=2 \delta t$) and the maximum scale was arbitrarily set to 4\,s. 

As the count rates derived from mask-weighting procedures
are already background subtracted, the pre-burst light curves
are dominated by a white Gaussian noise due to statistical fluctuations \citep{rizzuto07}.
In this particular case the wavelet coefficients are normally distributed \citep{l99}
and their power spectra follow a chi-square distribution 
with two degrees of freedom. Because of this property the significance levels 
of each peak in the wavelet power spectrum can be analytically derived \citep[see e.g.][]{tc98}.
We examined the global wavelet spectrum. i. e. the time-average over all the 
local wavelet spectra, for peak exceeding the noise spectrum level 
and set a minimum threshold of 99.7\% significance (corresponding to a 3$\sigma$ for a two-sided Gaussian distribution). 
In three cases (GRB~070406, GRB~080121 and GRB~091117) out of 49, 
the burst has been discovered in ground analysis and only $\sim$10 seconds of event data 
around the trigger time are available. Our search
has therefore been restricted to that interval. Because of border effects due
to the very short time interval, we did not apply the wavelet detection method 
and the light curves of these three bursts were simply inspected by eye.

We found that 4 out of 38 short bursts and 1 out of 11 bursts with extended
emission show a possible precursor activity.
Among them, one burst (GRB~090510) shows two precursors events, 
the former $\sim$13 s 
and  the latter, already known in literature \citep{abdo10}, 
$\sim$0.5 s before the GRB. 
The mask-weighted light curves of these GRBs are shown in Fig.~1. 
The vertical dashed lines in each panel mark the time 
interval of the candidate precursors.

With the selected threshold of significance (99.7\%) 
we expect a number of 0.04 spurious detections 
in each light curve, as we sampled 13 different time scales,  
and a total of $\sim$2 spurious detections in the whole sample.
Furthermore, we recall that the light curves derived using 
the mask-weighting technique are correct
if there are no other bright sources in the BAT field of view. 
While this assumption usually holds during the main GRB, 
when the source is weak, such in the case of precursor events, 
the contamination of nearby sources becomes significant and might therefore 
lead to spurious detections. Further analysis is therefore mandatory 
in order to verify if the features revealed by the wavelet algorithm 
are real and associated to the GRB. 

We proceeded by checking whether the 5 selected GRBs were detected by other
satellites and, if so, whether there was evidence of emission
simultaneously with the BAT candidate precursor. 
Three GRBs (GRB~081024A,GRB~090510 and GRB~09117) have this requisite and
in Fig.~1 we compare their \swift/BAT light curves
(reported in each upper panel) with the \fermi/GBM (GRB~090510 and GRB~081024A) 
and Suzaku/WAM (GRB~091117) light curves (reported in the bottom panel). 
Times are always given relative to the BAT trigger time. 
The cross-check of the light curves shows that the precursors
in GRB~081024A and GRB~091117 can be confidently considered real, 
as a simultaneous episode of emission has been observed by
{\it Fermi}/GBM and {\it Suzaku} respectively. 
Instead no significant emission above the background level
is observed in correspondence of the first precursor at $T_0$-13\,s 
in GRB~090510,
while the second precursor at $T_0$-0.5\,s is clearly detected by the {\it Fermi}/GBM.
GRB~090510 also triggered {\it Suzaku} and Konus-Wind, 
unfortunately no time-resolved events are available during 
the interval of the first precursor at $T_0-13$\,s and
any short time scale variability is hard to detect. 
Indeed Suzaku and Konus-Wind light curves (with a resolution of 1 s and 2.9 s respectively) do not show any significant excess at such early times
(K. Yamaoka, V. Pal'shin; private communications). 
This non-detection does not necessarily imply that the feature is spurious. 
Possible explanations are the smaller effective areas compared to BAT, or 
a precursor with a soft spectrum, e. g. peaking in the BAT energy range, 
as also expected on theoretical grounds.

\subsection{Imaging analysis}\label{sec:images}


\begin{deluxetable*}{lccccc}
\tabletypesize{\scriptsize} \tablecaption{Image significance of the candidate precursors.}
\tablewidth{0pt}
\tablehead{ \colhead{GRB\ \ \ \ \ \ \ \ \ \ \ \ \ \ \ \ \ \ \ \ \ \ } & \colhead{T$_i$} &  \colhead{T$_f$}  &  \colhead{Significance} & Probability\tablenotemark{$a$} & Others \\
             \colhead{ } & \colhead{[s]} & {[s]} & {[$\sigma$]} & { } & \colhead{ }  } 
\startdata
050724 (EE)\dotfill 	&  -108.5 	& -107.5 & 3.7 & 5\ee{-4} & -- \\
080702A\dotfill 	&  -140.6 	& -139.5 & 3.2 & 3\ee{-3} & -- \\
081024A\dotfill 	&  -1.70        & -1.45   & 5.5 & $<$\e{-5} & Fermi \\
090510 \dotfill 	&  -13.0 	& -12.6  & 5.2 & $<$\e{-5} & --\\
		        &  -0.55 	& -0.5   & 4.6 & \e{-5} & Fermi \\
091117\dotfill 	        &  -2.75 	& -2.65   & 1.8 & 6\ee{-2} & Suzaku

\enddata
\tablenotetext{1}{Probability of a spurious detection with equal or higher significance. Derived from Montecarlo simulations.}
\label{tab:image}
\end{deluxetable*}


In order to further check whether the excess in the light curve is related
to the GRB, we produced a background-subtracted sky image in the interval 
of the candidate precursor and searched for a source at the GRB position. 
This step allows a better characterization of the background level,
as the contribution of other nearby sources is properly removed.
Results are reported in Table~\ref{tab:image}, which lists the GRB name,
the precursor time interval, and the significance of the source in the image domain
as calculated by the tool \texttt{batcelldetect}.
If the precursor has been detected by other instruments (see \S~\ref{sec:wavelet}), 
they are listed in the last column. 

For a blind source detection a significance threshold of 6.5~$\sigma$ is usually adopted 
to confidently assess that a source is real. Indeed the Subthreshold
experiment\footnote{ \href{http://gcn.gsfc.nasa.gov/subthreshold.html}
{http://gcn.gsfc.nasa.gov/subthreshold.html}} carried out by the \swift~team 
showed that lowering this threshold
significantly increases the numbers of false detections ($\sim$96\% of false positives). 
However our search was not performed on the whole image,
as the source position was $a~priori$ known.  
This reduces the number of trials by a factor of $\sim$3\ee{4}, 
i.e. the number of independent pixels in a BAT image, 
with respect to a blind search
and the 6.5~$\sigma$ threshold poses therefore a too restrictive cut. 

We determined the probability to have a spurious N\,$\sigma$ detection 
at a fixed position through Montecarlo simulations. 
An inspection of the detector plane images (DPIs) shows no noisy detectors during the selected time intervals,
and therefore  statistical fluctuations are the dominant source of noise. 
By assuming a Poissonian distribution with a mean count rate of $\sim$0.12 cts\,s$^{-1}\,$det$^{-1}$, we simulated \e{5} source-free DPIs and derived the corresponding sky images. 
On each simulated image 
we ran the tool \texttt{batcelldetect} searching for a source at the 
GRB position. The probability that the detected source is due to background fluctuations is then calculated as
the ratio between the total number 
of fictitious detections with significance equal or greater than that of the precursor (Tab.~1, column 4) and the number of simulated images. The resulting values are listed in Tab.~1 (column 5).

\section{Results}\label{sec:res}

The results of our analysis are summarized in Tab.~\ref{tab:image}.
We found evidence of possible precursor activity 
in 4 short GRBs, out of a sample of 38 events, 
and only in 1 GRBs with extended emission (EE), 
out of a sample of 11 events.
One burst (GRB~090510) shows two precursors,
at $\sim T_0-$13 s and $\sim T_0-$0.5 s respectively.
The $\gamma$-ray light curves of these bursts 
have been shown in Fig.~1.
Our definition of precursor, detailed in \S~\ref{sec:data},
differs from those given in previous systematic studies \citep[e.g][]{k95,l05}, 
as it does not impose any particular constraint 
on the quiescence time or the instrumental trigger \citep[e.g.][]{b08}.
In only one case the classification of the event as precursor 
depends on our operational definition:
the latter precursor in GRB~090510 does not satisfy either the conditions of \citet{k95}, 
because of its short delay time from the main GRB, 
and of \citet{l05}, as the precursor event triggered the {\it Fermi}/GBM.

Because our wavelet analysis (\S~\ref{sec:wavelet})
has been carried out on a large
sample of events (46 GRB light curves) 
and probed various time scales (13, from 256\,ms to 4\,s) 
in each light curve, 
we expect the resulting sample of precursors
to be contaminated by $\sim$2 spurious detections.
A cross-check between \swift~and other satellites
confirms that at least three of our candidate precursors
are real (Tab.~1, col. 5), namely  the cases of GRB 081024A, 
GRB090510 (2nd precursor), and GRB 091117.

We verified that the detected excess in the light curve
corresponds to a point source at the GRB position in the image domain (\S~\ref{sec:images}; Tab~1, col. 3-4). 
The former precursor in GRB~090510 is detected at a $>$5\,$\sigma$ 
significance. Montecarlo simulations showed that 
the probability of being a background fluctuation is very low ($<$\e{-5}), 
in agreement with the high significance of the detection.
Two cases remain controversial. The precursors in GRB~080702A and GRB~050724
are detected at a significance of 3.2 and 3.7 $\sigma$ respectively, 
having a $\approx$\e{-3} probability of being spurious. They are also not been seen by other instruments. These two precursors
are very intriguing, as they show the longest delay times from the GRB triggers ($\gtrsim$100\,s) similar to those observed in some long GRBs. 
However, in the present study we are unable to confidently determine whether they are real features or not. 
In this context, it is worth noting that the only short GRB (BATSE trigger 2614) in the sample of \citet{k95} shows a precursor $\sim$75\,s before the main burst. This strengthens the idea that long delay times are possible in short GRB precursors, as we will discuss in \S~\ref{crack}.
\citet{b09} similarly proposed that a few short GRBs
in the BATSE sample are preceded by precursors, 
sometimes with very long delays. 
A cross-check between the results of \citet{b09} and the
BATSE 4B Catalogue \citep{batse4b} shows several incongruences. In particular, the durations quoted in the BATSE Catalogue are greater than 5 s.

\begin{figure}[!t]
\centering
\includegraphics[angle=270,scale=0.6]{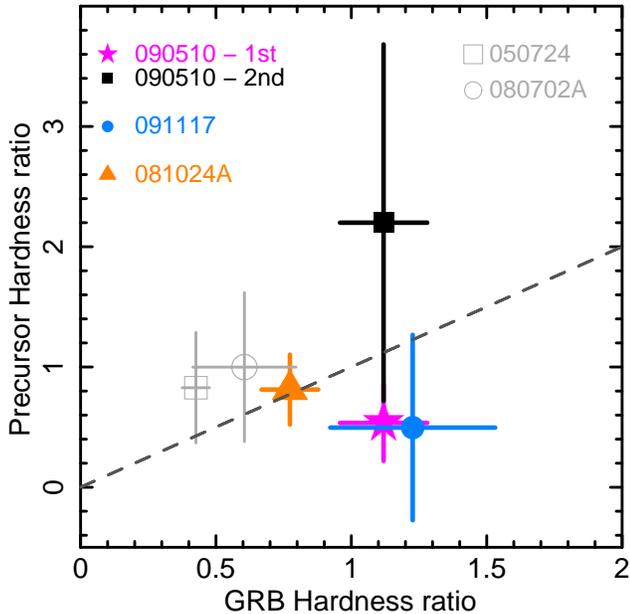}

\caption{Hardness ratio of the precursor vs. hardness ratio of the 
main GRB event. Error bars are at 1\,$\sigma$ confidence level.
The dashed line shows where the two ratios are equal.
The hardness ratio is defined as the count rate in the 50-150 keV band 
over the count rate in the 15-50 keV band.
We also report the precursors of GRB~050724 and GRB~080702A, albeit
we are unable to confidently determine whether they are real features, 
as explained in the text. }
\label{fig:hr}
\end{figure}

As our precursors are too faint to characterize their spectral shape,
we investigated the presence of a possible spectral difference
by comparing the precursor hardness ratio (HR) to that of the main GRB event.
This is shown in Fig.~\ref{fig:hr}: all the precursors appear consistent
with the main GRB properties, in agreement with the findings of \citet{b08}.
This result however might be partially a consequence of the \swift/BAT narrow bandpass.
For instance, the broadband {\it Fermi} light curves of GRB~090510 
(see \citealt{abdo10}, Fig.~1)
clearly shows that the main GRB event has an extremely hard spectrum,
peaking in the MeV range, while the precursor at $T_0 - 0.5$\,s peaks at around
200-300 keV. The precursor at $T_0 - 13$\,s, found in the \swift/BAT light curves, is even softer, peaking in the 15-50 keV energy band (Fig.~2, star symbol). 
On the other side, most theoretical models predict the peak of the precursor
emission in the X-ray energy range, i.e. at the lower end of the 
BAT energy threshold. Therefore, independently of the BAT bandpass, 
this should be reflected in our Fig.~\ref{fig:hr} by precursors occupying the region
with HR$\lesssim$1. Indeed the points seem to follow this trend.

We further investigated whether short GRBs with precursors differ 
from the other short GRBs in the sample, either in the prompt or
afterglow emission. We compared the distributions of their observed 
properties, such as $\gamma$-ray fluence, duration (T$_{90}$) and 
afterglow brightness, and ran a Kolmogorov-Smirnov (KS) test between the
the two samples (short bursts with and without precursors). 
The probability that they belong to the same GRB population is 36\%, 
based on the distribution of their fluences (in the 15-150 keV band),
and 34\%, based on the distribution of their durations. 
Similarly a comparison of the X-ray (0.3-10\,keV) afterglow fluxes distributions, 
observed at 100\, s and 1000\,s, shows no substantial difference 
(KS test probability of 96\% and 68\% respectively).

\section{Discussion}\label{sec:discuss}

Precursor activity has been so far associated to long GRBs.
Previous systematic studies have in fact been 
focused on the class of long GRBs, such as in the case of \citet{l05}  who excluded those bursts with a duration T$_{90}$$\leq$5 s, 
or biased against the detection of  short duration precursors by the low time resolution, such as in the case of \citet{k95}.
This has led to the common notion that precursors 
are not present in short GRBs.
For instance, \citet{mcbreen08} pointed out the presence
of a precursor at $T_0$-8\,s in the SN-less burst GRB~060505,
considering this as a further dissimilarity with the class of short GRBs. 
Our analysis showed instead that short GRBs
are also preceded by a precursor event, 
though less frequently than long GRBs (10\% vs. 20\% of long GRBs).
The precursors in our sample are charaterized by
short durations, never exceeding the GRB T$_{90}$. 
This disfavors the sub-jets model of \citet{nakamura00},
according to which the precursor duration is longer than
that of the main burst.

Only one burst with extended emission (GRB~050724) shows a possible  precursor $\sim$100\,s before the onset of the main GRB. However, as discussed in \S~\ref{sec:res}, we can not confirm in the present study whether it is a real event and therefore whether precursors are also present in bursts with EE.
It has been suggested that bursts with EE may be originated by a different progenitor system \citep{troja08, metzger08}. The current sample of short bursts with EE is still too small to draw any conclusion, but the absence of precursors in this subset of bursts, if confirmed by future observations, could provide a further evidence of their different nature.

The association between precursors and long GRBs has also driven 
most of the theoretical work, which has often related the precursor to the interaction of the jet with the massive star progenitor \citep[e.g.][]{err02,lb05}.
The presence of precursors in either long and short GRBs
might represent a challenge for such interpretation. 
Given the fact that in the internal shock model \citep{remes94, piran99} the GRB production is rather decoupled from the details of the central engine and both short and long bursts exhibit precursors, one my speculate that the precursor production is related to the fireball rather than the central engine itself. An obvious idea to test is whether the precursor could be caused by a fireball becoming optically thin \citep[e.\,g.][]{paczynski86} prior to the production of the prompt GRB emission. If we consider an ``isolated'' fireball (i.e. on becoming transparent the photons are not released into a surrounding, possibly intransparent environment), an observed duration of the precursor $\Delta t$ would imply a fireball radius (at this stage) of $R \sim 2 \Gamma^2 c \Delta t$,
where $\Gamma$ is the fireball bulk Lorentz factor. If we assume a saturated fireball with $\Gamma \approx \eta\equiv E/M c^2$, where $E$ is the fireball energy and $M$ its baryonic mass loading, and equate the above radius to the one where the fireball should become transparent to its own photons \citep{abramo91,piran99}, we find a relation between $\Gamma$ and $E$ for an observed duration $\Delta t$:
\begin{equation}
\Gamma \approx 25 \; E_{51}^{1/5} \left( \frac{1 {\rm s}}{\Delta t} \right)^{2/5}
\end{equation} 

 If, as indicated by recent {\it Fermi} results \citep{fermi10}, short GRBs do indeed possess Lorentz factors in excess of $10^2$, a precursor origin related to a fireball becoming optically thin, would require a large fireball energy, $E > 10^{53}$ erg.
 
Moreover, if we assume that the main GRB signal is produced by internal shocks, then the variability time scale $\delta t_{\rm var}$ can be restricted to $\delta t_{\rm var} \gtrsim \Delta T$ \citep{l05}, where $\Delta T$ is the
time interval between the precursor and the main prompt emission. The observed delays would suggest implausibly long variability time scales, longer than the main GRB duration itself. 
Thus, at least if the prompt emission is caused by internal shocks, we consider it unlikely that the observed precursors are produced by fireballs becoming optically thin. This conclusion could need to be modified if the fireball is released into an optically thick surrounding, say from a previously ejected wind (see \S~\ref{wind}).

\subsection{Central engine-related mechanisms}\label{sec:parent}
Mergers of compact binaries, either in the form of a double neutron star (DNS) 
\citep{blinnikov84,paczynski86,goodman86,eichler89} or a neutron star-black hole
system (NS-BH)\citep{paczynski91,narayan92}, are still arguably the most likely
central engines of short GRBs. In the following we will focus on how such systems
may produce an electromagnetic transient prior to the main GRB.

\subsubsection{Interaction of neutron star magnetospheres }
\citet{hansen01} model the electromagnetic signatures
that result from the interaction of two NS magnetospheres prior 
to a double neutron star merger. The main prediction of their model 
is an X-ray transient preceding the merger on a time scale of a few seconds.  
This signal could also be accompanied by a radio pulse.

\citet{hansen01} consider a binary system consisting of an old, recycled pulsar
that is rapidly spinning ($P \approx 1 - 100$ ms) and possesses a magnetic field
of moderate strength ($B \sim 10^9-10^{11}$ G) and a younger, slowly rotating 
($P \approx 10 - 1000$ s) strong-field ($B \sim 10^{12}-10^{15}$ G) neutron star 
(possibly a magnetar), a combination that can be expected on evolutionary grounds.
If the magnetar birth rate is about 10\% of the ``ordinary pulsar" birth rate, a decent fraction of double neutron stars should contain magnetars, at least initially.
The recycled pulsar is considered as a perfectly conducting sphere that passes
through the external field prescribed by the magnetar. In this way a dipolar 
magnetic field is induced whose magnetic dipole is directed against the external
magnetic field. The motion of the pulsar through the external field induces 
surface charges that in turn produce electric fields with a component along
the total magnetic field which accelerate charges in an attempt to short out
this parallel electric field component. Once energetic enough, the latter
produce curvature photons together with a dense population of electron-positron 
pairs. Pair plasma released into regions of increasing magnetic field strength
are likely to be trapped in a optically thick cloud, while those released into
regions of decreasing field strength result in a relativistically expanding
wind of pairs and photons.

The strongest prediction of this model is the presence of an early precursor 
produced by the relativistic wind.
The precursor spectrum should be close to thermal and hardening 
as the stars are driven towards coalescence. Interestingly, GRB~090510 has two precursor signals, 
where the first one peaks in the 15-50 keV energy band while the second peaks around 300 keV. 
Such behavior would be consistent with the predictions of the Hansen-Lyutikov model.
The maximum luminosity that the precursor can reach is of the order of: 
\begin{equation}
L \approx 7 \times 10^{45} {\rm erg}\,{\rm s}^{-1} \left(\frac{B}{10^{15} {\rm G}}\right)^2 \left(\frac{a}{10^{7}\,{\rm cm}}\right)^{-7}
\end{equation}
It follows that in order to match with the observed properties of short GRB precursors 
a NS with a magnetar-like field ($B>$\e{15}\,G) is required.
Such strong magnetic fields likely decay on much shorter time scales ($\sim$\e{4}-\e{5} yrs; \citealt[and references therein]{hk98,harding06}) than the merger lifetime,
and a NS with a moderate magnetic field $B\sim$\e{12}-\e{13} G looks a more plausible configuration. Some population synthesis models \citep{belc02,belc06} however predict that a sizable fraction
of DNS mergers has much shorter inspiral times. This short-lived channel peaks at an inspiral time of $\sim$3\ee{5} years and after this time the magnetic field should have decayed by only a factor of a few \citep[see Fig.~1 of][]{hk98}.

\subsubsection{Neutron star flares induced by tidal crust-cracking}\label{crack}
As a compact binary system secularly spirals in, the neutron star(s) become(s)
vulnerable to tidal distortion. At a separation $a$ the companion induces an 
ellipticity of $\epsilon_1 \sim \delta R_1/R_1 \sim \frac{m_2}{m_1} \left(\frac{R_{\rm ns}}{a}\right)^3$, 
where $m_1$ and $m_2$ are the NSs masses and $R_{\rm ns}$ is the NS radius.
Once the ellipticity exceeds a critical value, the neutron star crust cracks and likely triggers
a violent restructuring of the magnetic field that may go along with a reconnection flare, similar 
to what is thought to happen in a magnetar giant flare \citep{thompson95,palmer05,hurley05}.
Once the crust has been cracked for the first time, the neutron star enters a ``tidal grinding phase'' in which 
the tides exert a constant restructuring of crust, likely going along with further magnetic field 
reconfiguration and dissipation.

The exact numerical value of the critical ellipticity that the crust can still sustain, 
$\epsilon_{\rm c}$, is not well-known, but recent studies based molecular dynamics simulations 
\citep{horowitz09} suggest that neutron star crusts can sustain strains up to a breaking value of 
$\sigma_{\rm max}\approx 0.1$, corresponding to critical ellipticities up to $\epsilon_{\rm c} \approx 4 \times 10^{-6}$ 
\citep{ushomirsky00,owen05}.
Thus, the tidally-induced crust cracking is expected to occur at a separation of 
\begin{equation}
a_{\rm crit}\approx 100 \; \left(\frac{m_2}{m_1}\right)^{1/3} \epsilon_{{\rm c,}-6}^{-1/3} R_{\rm ns} 
\end{equation}
where $\epsilon_{{\rm c,}-6}^{-1/3}$ is the ellipticity in units of $10^{-6}$. Applying the point-mass
limit for a circular binary system (ignoring the effects of the finite stellar radii; \citealt{peters64}) one finds
for the duration of the tidal grinding phase prior to the merger:
\begin{eqnarray}
\tau_{\rm tg}&\approx&\frac{5}{256} \frac{c^5}{G^3} \frac{a^4_{\rm crit}}{m_1 m_2 (m_1+m_2)} \nonumber\\
&\approx& 62 \; {\rm min} \; \epsilon_{{\rm c,}-6}^{-4/3}   \frac{\; 2 q^{1/3}}{1+q}
\left(\frac{m_{\rm ns}}{1.4 \; {\rm M}_\odot}\right)^{-3} \left(\frac{R_{\rm ns}}{10\; {\rm km}}\right)^4 
\label{eq:ttg0}
\end{eqnarray} 
where we defined the mass ratio $q=m_2/m_1$.
Thus for a binary system with the most likely parameters one expects the crust restructuring to set in about an hour
ahead of the burst (where we have assumed the delay between coalescence and burst is negligible). Should the crust be able to sustain substantially larger deformations, say an order of magnitude more, this duration could be brought 
down to minutes.
Due to the larger total mass the tidal grinding duration for NS-BH binaries 
is somewhat shorter than the above estimate, but for the low mass black holes that are most interesting for
GRBs, e.g. \citet{rosswog05a}, the difference is just a factor of two. 

Naively, one would expect the major flaring activity to occur coincident with the first crust cracking (Eq.~\ref{eq:ttg0}),
and just by analogy with magnetar giant flares \citep{palmer05}, such precursors should have spectral properties similar to the
properties of the main burst. 
The elastic energy stored within the deformed NS crust is $\sim$10$^{46}$\,($\sigma_{\rm max}$/0.1)$^2$\,erg~\citep{thompson01}, and if this is the main
energy source, the corresponding precursors would not be visible
beyond 40-80 Mpc.

\subsubsection {A relativistic jet ploughing through a pre-ejected, neutrino-driven baryonic wind} \label{wind}

Directly after the merger --but possibly before a relativistic jet can be launched--
the remnant of a neutron star merger consists of a hot, differentially rotating, super-massive neutron
star, surrounded by a massive ($\sim 0.1$ M$_\odot$), thick accretion disk of 
neutron-rich debris \citep[e.g.][]{ruffert01,rosswog03c}. In the inner parts of this
disk, at a distance $r$ from the centre of the central object, a nucleon is 
gravitationally bound with an energy of $E_{\rm grav} \approx 35 \; {\rm MeV} 
\left(\frac{M_{\rm co}}{2.5 \; {\rm M}_\odot}\right) \left(\frac{100 \; {\rm km}}
{r}\right)$. On the other hand, a large fraction of gravitational binding energy 
of the binary system is released in the form of neutrinos, with average energies
of $\langle E_{\nu_e}\rangle \approx 10$ MeV,  $\langle E_{\bar{\nu}_e}\rangle 
\approx 15$ MeV and $\langle E_{\nu_X}\rangle \approx 20$ MeV 
\citep{ruffert01,rosswog03a} where the index $X$ refers collectively to the heavy 
lepton neutrinos. It had been realized early on  \citep{ruffert97a,rosswog02b}
that such a configuration should ablate a substantial fraction of the debris in 
a neutrino-driven, baryonic wind. A quantitative calculation beyond order of
magnitude estimates has only recently become possible \citep{dessart09}.
This study found that a bi-polar, non-relativistic ($v \sim 0.1 c$) wind of 
$\dot{M} \sim 10^{-3}$ M$_\odot/s$ heavily pollutes the polar regions and prevents the 
formation of ultra-relativistic outflow. Thus, for at least as long as the central 
object has not collapsed into a black hole, it seems impossible to produce a GRB.

After a (likely, but not necessarily guaranteed) collapse, one is left with the 
``standard'' central engine, a black hole-disk system.  Once this happens, 
the neutrino-driven mass loss will be seriously reduced and jet formation seems likely.
It may be speculated that the emerging relativistic jet has to plough through 
the pre-ejected neutrino-driven baryonic cloud, possibly producing a precursor signal.
The details of such a jet-cloud interaction are very likely rather involved and a
quantitative investigation of this issue is beyond the scope of this paper.
We note however that as in this case the precursor marks the start of the central engine activity,
the merging process could happen significantly before the observed main burst, 
stretching the temporal window over which a gravitational waves signal must be searched \citep{ligo10}.

\subsection{Constraints on quantum gravity}\label{sec:liv}

GRB~090510 is accompanied by high-energy ($>$100 MeV) emission, lasting up to 200 s
after the burst \citep{agile10,max10}. The detection of very energetic photons (up to 31 GeV) during
the main prompt emission and the high redshift of the source ($z$=0.903; \citealt{rau09}) 
led to very tight constraints on the quantum gravity mass M$_{QG}$, 
excluding a possible linear energy dependence 
of the propagation speed of light \citep{abdo10}.
The authors adopt two different approaches to constrain Lorentz invariance
violation effects: the former, based on the method outlined in \citet{scargle08}, derives a limit on the quantum gravity
mass of M$_{QG}$$>$1.22 M$_{Pl}$, where M$_{Pl}$=1.2\ee{19}~GeV/c$^2$ is the Planck mass. This limit remains unchanged by the present findings. 
 
As the photon emission time and location are unknown, 
the latter approach conservatively assumes that the observed 31 GeV 
photon has not been emitted before the onset of the low energy emission. 
Under this assumption, the limit on the quantum gravity
mass is M$_{QG}$$>$1.19 M$_{Pl}$. 
In their calculation \citet{abdo10} considered the precursor at T$_0$-0.5 s, 
also detected by the \fermi/GBM (see Fig.~\ref{fig:batlc}), as the earliest possible emission time. 
The detection of an earlier precursor, presented in this work, 
shows that emission started well before the GRB,
implying a maximum delay of $\sim$13.3 s between the lowest and highest energy photons.  
The corresponding  upper limit on the quantum gravity mass is therefore
significantly reduced to  M$_{QG}$$>$0.09\,M$_{Pl}$.

\section{Conclusion}

We carried out a systematic search of precursors on the sample
of short GRBs observed by \swift. 
We found that $\sim$8-10\% of short GRBs
shows such early episode of emission, preceding
the main GRB by a few seconds ($\Delta T$$\leq$13\,s). 
In our sample we found some evidence, though not yet conclusive,
that the observed delay can be as long as $\sim$100\,s. 
The spectral properties of these precursors do not substantially 
differ from the prompt emission. This result however might be partially 
a consequence of the \swift/BAT narrow bandpass.

We consider it unlikely that the observed precursors are 
produced by fireballs becoming optically thin, and argue
instead that the preburst activity in short GRBs
is related to their progenitors, i. e. compact objects mergers.
We discuss three possible central engine-related mechanisms:
the interaction of neutron star magnetospheres \citep{hansen01},
which requires one of the two compact objects to be a magnetar;
flares from NS crust cracking, which predicts very long delays
between the precursor and the main burst. 
Finally, analogously to long GRB precursors which are associated 
to the interaction of the relativistic jet with the stellar envelope, 
the precursors in short GRBs might be produced by the jet 
interacting with a pre-ejected neutrino-driven baryonic wind. 
In this last case, the precursor is produced after the merger 
and marks the start of the central engine activity.


\acknowledgements{}

We thank G. Skinner and C. Markwardt for discussions and 
useful suggestions on the \swift/BAT data analysis. 
This research was supported by an appointment to the NASA
Postdoctoral Program at the Goddard Space Flight Center, 
administered by Oak Ridge Associated Universities 
through a contract with NASA.


\bibliographystyle{aa}
\bibliography{precursor}

\end{document}